# Tuning the low-energy band structure in twisted bilayer $WSe_2$

T.-H.-Y. Vu[1,*], O. J. Clark[1,*], N. H. Jo[2], J. Blyth[1], Q. Li[1], C. Jozwiak[2], A. Bostwick[2], J. B. Muir[3,4], L. Jia[3,4], J. A. Davis[3,4], I. Di Bernardo[1,5], A. Grubišić-Čabo[6], K. Xing[1], W. Zhao[7,8], S. H. Ryu[2], S. H. Lee[9,10], Z. Mao[9,10], K. Watanabe[11], T. Taniguchi[12], B. A. Chambers[13], S. L. Harmer[13,14], E. Rotenberg[2], M. S. Fuhrer[1,8], M. T. Edmonds[1,8,15,†]

[1]School of Physics and Astronomy, Monash University, Clayton, VIC, Australia

[2]Advanced Light Source, Lawrence Berkeley National Laboratory, Berkeley, CA, 94720 USA

[3]Optical Sciences Centre, Swinburne University of Technology, Hawthorn, 3122, Victoria, Australia

[4]ARC Centre of Excellence in Future Low-Energy Electronics Technologies, Swinburne University of Technology, Hawthorn, 3122, Victoria, Australia

[5]Departamento de Física de la Materia Condensada, Universidad Autónoma de Madrid, Madrid, 28049 Spain

[6]Zernike Institute for Advanced Materials, University of Groningen, Nijenborgh 3, 9747 AG Groningen, The Netherlands

[7]Department of Materials Science and Engineering, Monash University, Clayton VIC 3800, Australia

[8]ARC Centre for Future Low Energy Electronics Technologies, Monash University, Clayton, VIC, Australia

[9]Department of Physics, Pennsylvania State University, University Park, PA, 16802, USA

[10]2D Crystal Consortium, Materials Research Institute, Pennsylvania State University, University Park, PA, 16802, USA

[11]Research Center for Electronic and Optical Materials, National Institute for Materials Science, 1-1 Namiki, Tsukuba 305-0044, Japan

[12]Research Center for Materials Nanoarchitectonics, National Institute for Materials Science, 1-1 Namiki, Tsukuba 305-0044, Japan

[13]Flinders Microscopy and Microanalysis, Flinders University, Bedford Park, South Australia 5042, Australia

[14]Institute for Nanoscale Science and Technology, Flinders University, Bedford Park, South Australia 5042, Australia

[15]ANFF-VIC Technology Fellow, Melbourne Centre for Nanofabrication, Victorian Node of the Australian National Fabrication Facility, Clayton, VIC 3168, Australia

* Equal contribution

† Corresponding author

**ABSTRACT**. Tuning the electronic structures of two-dimensional (2D) material-based heterostructures is of crucial importance for their use in functional next-generation electronics. Here, through angle-resolved photoemission spectroscopy with nanoscale spatial resolution (nano-ARPES), we systematically track the evolution of the near-Fermi-level electronic structure of bilayer $WSe_2$ over a large range of twist angle. While the momentum positioning of the valence band maxima is independent of twist angle, we find that the energetic separation between the hole bands at the K point of the Brillouin zone and the higher binding-energy hole band at Γ can be varied in excess of 100 meV. We explore the mechanisms underpinning this evolution and discuss the implications for tuning both the size of the band gaps, and the efficiency of the spin-dependent electron-phonon coupling channels in homobilayer transition-metal dichalcogenide devices.

**Keywords:** Transition-metal dichalcogenides, twistronics, nano-angle resolved photoemission spectroscopy, twisted heterostructures, 2D semiconductors

## I. INTRODUCTION.

The creation of artificial heterostructures from lattice-mismatched monolayers of van der Waals materials has unlocked a new dimension to the parameter space in which solid-state electronic properties can be engineered and controlled [1–6]. In particular, recent research has focused on invoking highly correlated electronic phases into bilayers of systems which otherwise are very weakly correlated. This follows the seminal work on "magic angle" twisted bilayer graphene, wherein unconventional superconductor and Mott-insulator behavior can be induced with an extremely precise control of the moiré potential, or equivalently, the twist angle between the monolayer sheets [7–14]. While remarkable, the precision in twist angle required to realize similar physics in other heterobilayer systems, e.g. in the 2H structured transition-metal dichalcogenides (TMDs) [8–14], has led to an imbalance in the exploration of this new phase space, with most studies focusing on the small-angle regimes in which correlated phases are most likely to occur [15].

In this study, we aim to explore broader-scale electronic structure variations over a complete angular range. We focus on 2H-$WSe_2$ which, when commensurately stacked, undergoes an indirect-to-direct band-gap transition with reducing thickness due to the discretisation of the highly three-dimensional Γ-point hole bands, thus leaving the VBM at the K points [16–20]. In the monolayer limit, this unlocks enormous potential for 'spintronic' applications due to the valley-locked spin polarization which can be selectively excited with circularly polarised light sources, therefore enabling spin-polarized current generation from monolayer TMD devices [5,16,21–25].

A similar dependency on the relative energetic positioning of the Γ and K-point valence band maxima can be inferred across studies of twisted bilayer $WSe_2$ systems [1,26]. While both the positioning of the conduction band maximum and valence band maximum are expected to remain unchanged for all twist angles, the ability to tune the width of the resulting band gaps through the manipulation of the shallowest-energy valence bands could greatly affect potential transport properties in gated systems [1,14], including, for example, the available spin-dependent electron-phonon scattering channels [27].

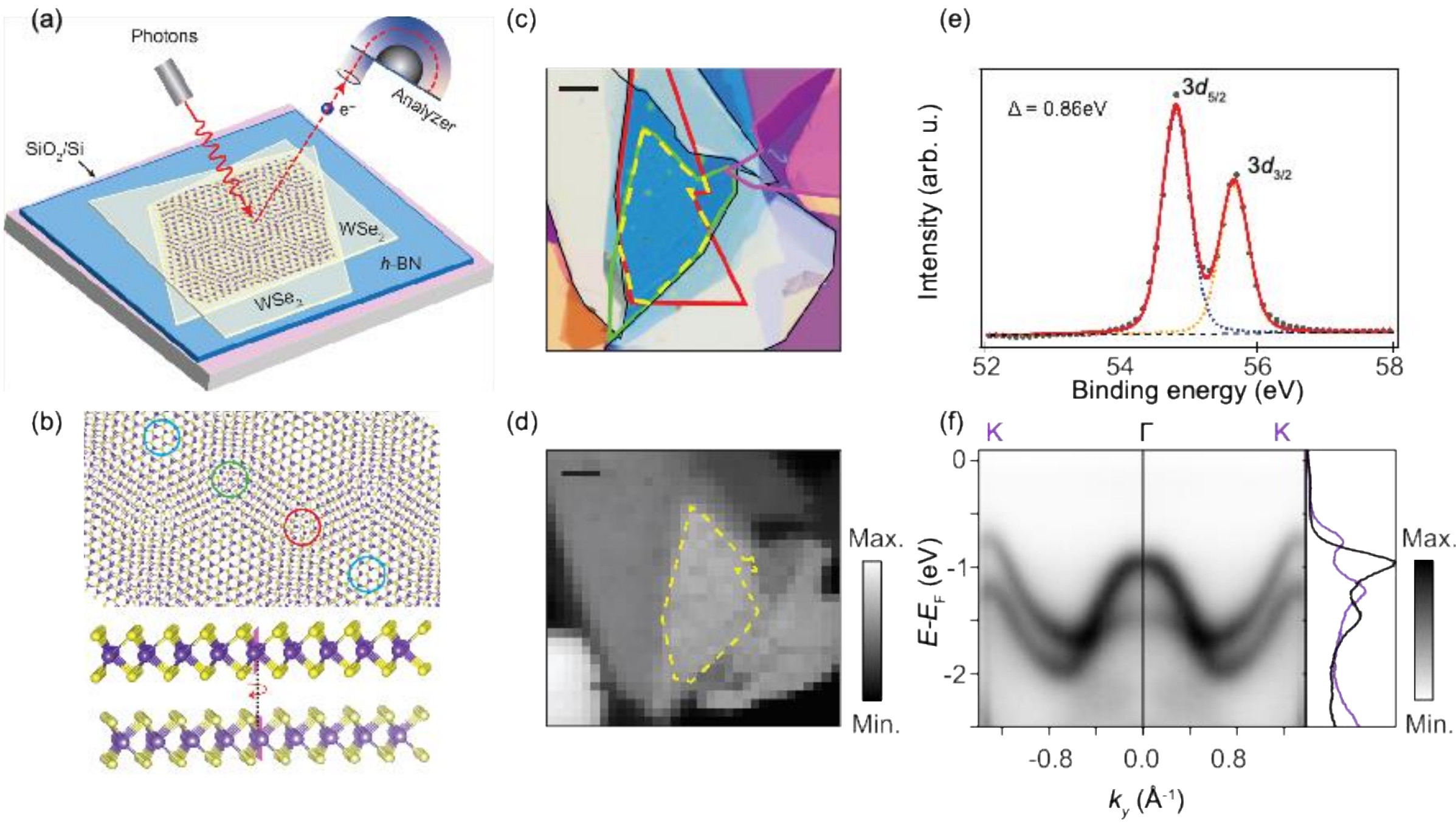

Figure 1 - (a) Schematic nano-ARPES measurements on the twisted bilayer $WSe_2$ device. (b) Schematic of an exemplary moiré superlattice formed in twisted bilayer $WSe_2$ at a nominal twist angle of 14°. The blue, red and green circles highlight the AA, BA and AB regions, respectively. (c) Optical microscope image of the twisted region. Top and bottom monolayers are highlighted by green and red lines, respectively. Bulk TMD regions are highlight by black lines and the twisted bilayer region is highlighted by the yellow dashed line. (d) Spatial valence band mapping of the same region in c). The scale bar in c and d is 10 μm. (e) Se 3*d* core-level spectra (hν=145 eV) of the sample shown in c. (f) Twisted bilayer band dispersion (symmetrized about Γ) and energy distribution curves (EDCs) at Γ (black) and K (purple). The twist angle is 14°.

Here, through angle-resolved photoemission spectroscopy with nanoscale spatial resolution (nano-ARPES), we characterise the origin and scale of these band shifts, showing how the twist-angle dependent interlayer hopping channels affect the functional properties of the electronic structure, even without the previously observed correlated phases emerging from moiré-induced "flat bands".

## II. METHODS.

Twisted bilayers of the TMD $WSe_2$ were fabricated using standard dry-transfer stacking methods. The fabrication process begins with using silicone-free blue tape to exfoliate hexagonal boron nitride (*h*BN) onto a 285-nm conductive silicon/silicon dioxide wafer before flat flakes of *h*BN with suitable thickness were selected. Then the $WSe_2$ monolayers were prepared on polydimethylsiloxane (PDMS) substrates using the same blue tape exfoliation method. Monolayers were confirmed using both optical contrast and photoluminescence spectroscopy techniques [28], then their orientation was determined by second harmonic generation spectroscopy [29]. The $WSe_2$ monolayers were then transferred onto the *h*BN one by one by dry-transfer method with desired twist angle to obtain twisted bilayer $WSe_2$. $WSe_2$ flakes are grounded via graphene to Au contacts in order to prevent charging during photoemission measurements. After each transfer, the sample was cleaned by diisopropylamine, followed by annealing in an argon environment at 100°C for 1 hour. ARPES measurements were carried out using *p*-polarised synchrotron light of energies between 24 and 150 eV at the Beamline 7.0.2 (MAESTRO) at the Advanced Light Source (ALS). Photoelectrons were detected with an Omicron Scienta R4000 hemispherical electron analyzer, and the base pressure of the setups was better than $1\times10^{-10}$ Torr. Samples were cooled to measurement temperatures of approximately 10 K following 3+ hour UHV annealing at 270 °C.

## III. RESULTS.

Figure 1(a) schematizes moiré superlattices assembled from twisted layers of hexagonally structured 2D materials. The long-range moiré periodicity is determined by the combination of the relative twist angle between constituent layers, and the degree of lattice-constant mismatch in the cases where the two monolayers are inequivalent. Such heterostructures have their low-energy electronic structures significantly modified through the coupled influence of (1) Brillouin zone (BZ) folding due to the moiré, Fig. 1b, which enforces hybridization of states otherwise well separated in momentum space, and (2) interlayer hybridization enabled by the newly opened hopping channel normal to the surface plane. It is the first of these two effects that is largely responsible for the correlated phenomena observed in small-angle twisted bilayer graphene and other 2D materials, where flat bands appear at the BZ center, produced from the hybridization of bands from K and K' valleys [7,30–33]. The latter effect can be expected to be significant for all twist angles, with details of the interlayer hopping channels (significant for out-of-plane $d_{z^2}$-orbitals, for example) delicately dependent on the relative atomic species and orientations either side of the van der Waals gap. For nonzero twist angles, relative atomic alignments are spatially dependent, thereby producing a periodic modulation of interlayer coupling across the heterostructure surface. This spatial dependence is schematized in Fig. 1b, where non-zero twist angles result in laterally separated regions of AA and AB-stacked bilayers, manifesting in well-defined ferroelectric domains [34–37].

An example heterobilayer $WSe_2$ device is shown in Fig 1c. The spatially resolved valence band mapping over the heterostructures, Fig. 1d, replicates well the optical image, thus allowing high precision in the selection of the region of interest in photoemission data. The Se 3*d* core level spectrum, Fig. 1e, shows clean, sharp peaks demonstrating a lack of oxidation.

In Fig. 1f, a representative Γ-K band dispersion is shown for a 14° twisted device. Energy-distribution curves (EDCs) are shown in the inset for the Γ and K points. There are two bands at each of the Γ and K points. The pair of hole pockets at K and K' originate from a Rashba-type spin-orbit splitting due to the in-plane electric dipole within each Se-W-Se unit [21–25,38]. Within each constituent monolayer, a large out-of-plane spin polarisation is therefore present, switching between the hole pockets, and from K to K' in accordance with time-reversal symmetry [38]. In AB stacked bilayers (0° twist angle), spin degeneracy is restored due to the opposite dipole orientation within the two sublayers. The spin polarisation can therefore only be unlocked with a breaking of layer degeneracy [39] or with a layer-selective excitation [38,40–43]. This is not the case for the twist angle range $0° < \theta < 120°$, where an inversion symmetry-breaking net dipole exists, producing the aforementioned ferroelectricity. In contrast, the hole bands at Γ are spin degenerate for both monolayer and bilayer systems, independent of the twist angle. The two bands are instead distinguished by the $k_z$ quantum number: The W $d_{z^2}$-orbitals from which the bands derive have significant hopping along the c-axis, thus translating to a large continuous $k_z$ dispersion of the corresponding bands in the bulk system [38]. The bonding-antibonding hybridization gap between the two quantized states in the bilayer is therefore of similar magnitude to the bulk $k_z$ bandwidth. The second band at Γ is absent in monolayers due to the absence of $d_{z2}$-orbital overlap along the c-axis.

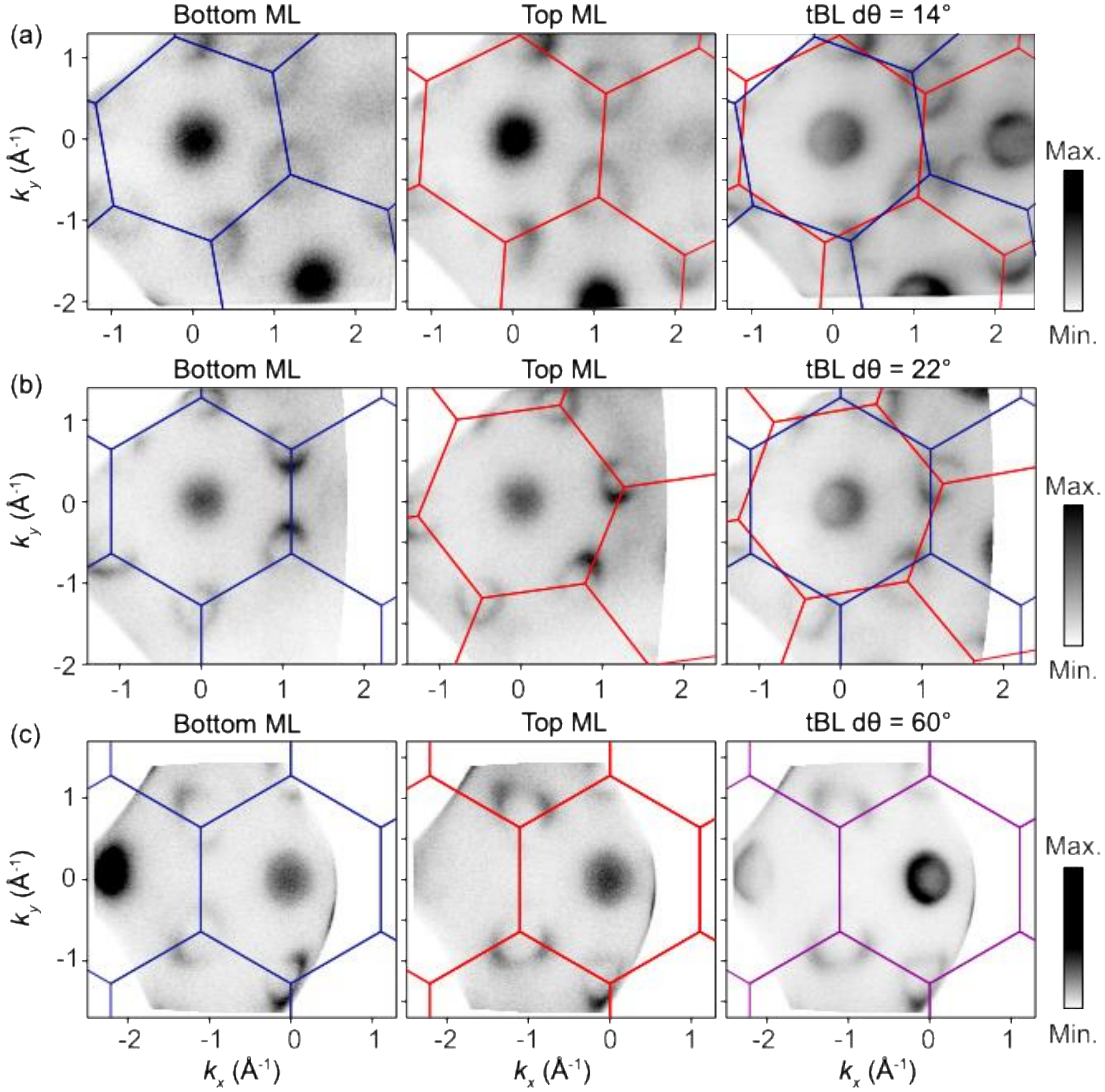

Figure 2 - Determining the relative twist angles for homo-bilayer $WSe_2$ devices. Bottom (left) and top (middle) monolayer constant energy contours are displayed for $E-E_F = \sim 1.25$eV and fine-tuned to optimise the matrix element symmetries. The corresponding constant energy contour for the twisted bilayer (tBL) regions are displayed (right), for (a) 14°, (b) 22° and (c) 60° rotated devices. The overlaid Brillouin zones contours use a lattice constant of a=3.28 Å [17]. We note the presence of extra bands deriving from graphene in a subset of these spectra. These derive from the graphene contacts used within the heterostructures (see Methods)

To accurately determine the relative twist angle for our twisted bilayer devices, representative photoemission constant energy $k_x$-$k_y$ contours are shown in Fig. 2 (a-c). Care is taken during device construction to preserve regions of isolated top and bottom monolayer regions to determine the twist angle *in situ*. This approach provides a second test of the twist angle following preliminary characterisation of individual flakes using second harmonic generation (SHG) [44]. This therefore negates any uncertainties caused by possible lattice relaxations during annealing processes. Moreover, one can distinguish between AB- (0°) and AA-stacked (60°) regions due to inequivalent $C_3$-symmetric photoemission matrix elements [45]. Figure 2 shows three such sets of Fermi surface maps approximately 1.25 eV below the valence band maxima. For 14° and 22° twisted bilayer devices, Fig. 2a and 2b, respectively, the distribution of spectral weight around the K point pockets is equivalent between the top and bottom monolayers, thus indicating a relative orientation closer to 0° than 60°. This is not the case for the set of constant energy contours shown in Fig. 2(c), in which there is a clear change in the photoemission matrix elements around the K pockets. The corresponding constant energy contours for the twisted bilayer regions are, to first approximation, a superposition of these two monolayers, with their relative intensity reflecting the finite electron mean free path in photoemission experiments [45]. Finer changes to the band structure due to the combined influence of interlayer coupling and moiré band renormalisation are explored in Figure 3.

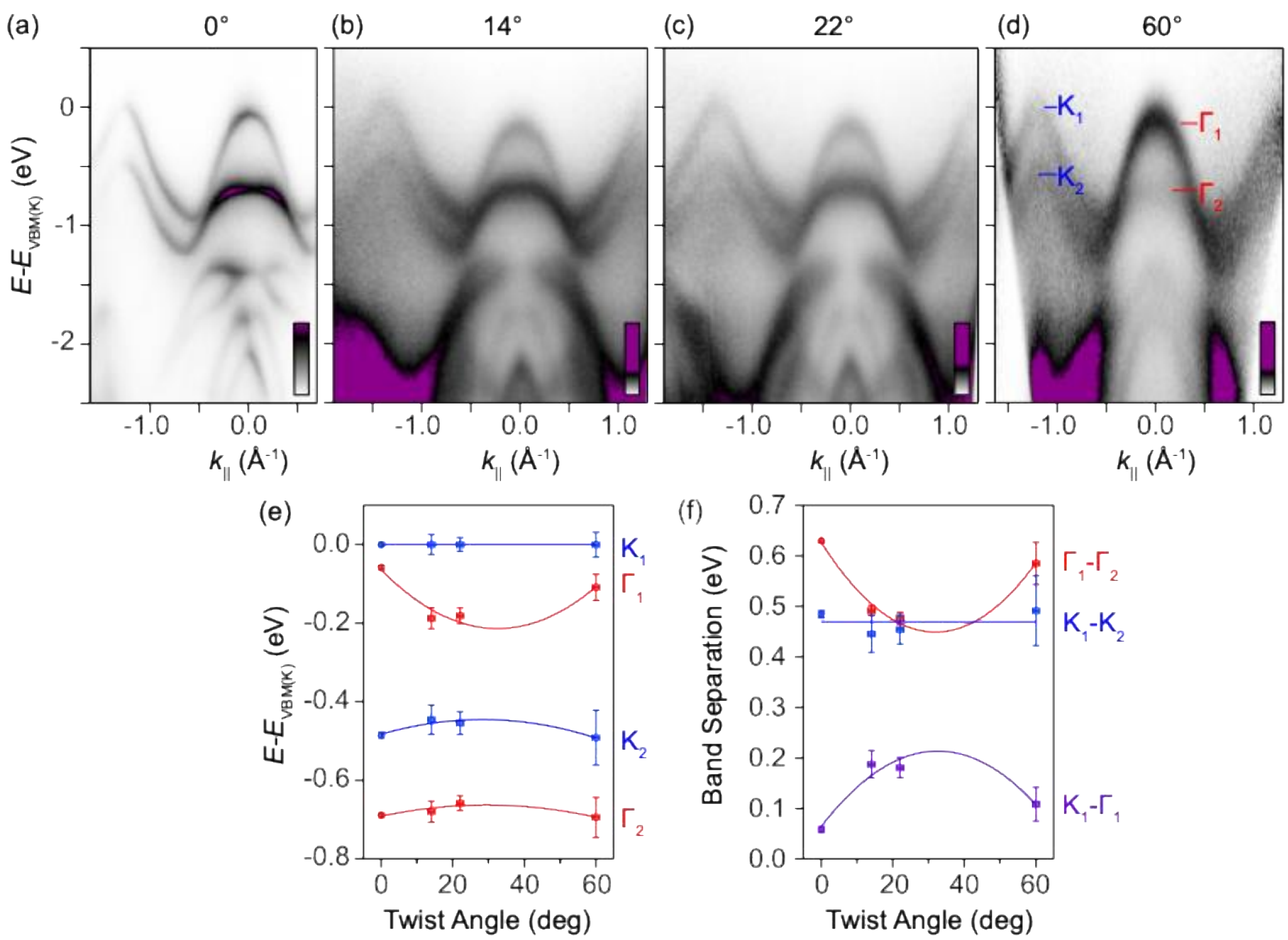

Figure 3 - (a-d) High symmetry ARPES K- Γ-K band dispersions for four representative twisted bilayer $WSe_2$ devices. The energy axis is displayed relative to the K point valence band maxima ($K_1$) to remove uncertainties surrounding Fermi level positioning. In (b-d), a normalisation across the image has been applied, to equalize the intensities of all energy distribution curves (EDCs), thereby increasing spectral weight of bands at the K points relative to those at the Γ point. Raw data are shown in the Supplementary Information. In (d), the four highest energy bands are labelled for comparison with (e-f). (e) Extracted band positions (see Methods) as a function of twist angle for the indicated bands. Solid lines are linear or parabolic trendlines for bands at the K and Γ points, and are intended as guides to the eye. (f) Band separations as a function of twist angle.

In Figure 3, high symmetry band dispersions capturing the Γ and K points, corresponding to the Fermi surfaces in Figure 2, are displayed as a function of twist angle. We note that, although the experimental conditions for these energy dispersions differ, the band positions should not be influenced by the switch of geometry or by a change of photon energy due to their two-dimensionality. The dependence with twist angle of the positioning of the shallowest four bands are displayed in Fig. 3(e-f). In all cases, positions are extracted from EDC fitting to Gaussian-broadened Lorentzians, accounting for finite experimental resolution and a Shirley background, shown in Supplemental Figure 1. Any observed trends should be strictly 60° periodic, reflecting the underlying $C_3$ symmetry of these systems and the equivalency in the energetics of the bands at the K and K' points.

The extracted peak positions of the four shallowest energy bands, and their relative separations are displayed in Fig. 3(e-f). Changes are tracked relative to the valence band maximum of band $K_1$ (see Fig. 3d). The relative separation of the pair of K point bands does not change within the experimental uncertainties. For twisted samples, the details of the available hopping channels are inconsequential for the in-plane d-orbitals from which the pair of K point bands derive. This is most readily seen when comparing monolayer, bilayer and bulk electronic structures [1, 26, 46, 47, 38], where the band structure at K remains unchanged. In the small-angle regime where the moiré potential is particularly strong, cross-BZ hybridization with bands with similar energetics is expected, most notably with the $d_{z^2}$-derived hole bands at the Γ point, which have significant c-axis hopping due to their extended spatial extent into the van der Waals gap. This could potentially induce an interlayer sensitivity to the K point bands for the angles in which cross-BZ hybridization is particularly favourable, though we do not see such effects here. As $K_1$ and $K_2$ are two-fold degenerate in bilayer systems, one may instead expect a fine structure to arise to a c-axis inversion asymmetry through the heterostructure, which is known to lift layer degeneracy in e.g. bulk $WSe_2$ samples where a surface potential is applied [39], or due to the creation of non-degenerate $k_z$ subbands due to hybridization with bands more susceptible to the available interlayer hopping channels. Such a splitting is not observed in the datasets in Figure 3, though we note the large intrinsic broadening enables only for the placement of a ~60 meV upper limit on a potential splitting. Together, $K_1$ and $K_2$ remain unchanged as a function of twist angle within experimental error.

In contrast, not only do the Γ-centered bands, $\Gamma_1$ and $\Gamma_2$, move relative to the K point VBM, but their separation also changes. $\Gamma_1$ moves to deeper binding energies with increasing twist angle, finding a turning point around the high-energy (i.e. maximally mismatched) 30° configuration, before relaxing back towards its initial configuration in the 60° orientation. $\Gamma_2$ is relatively static in comparison, though a small observed shift to shallower binding energies is observed within the experimental uncertainties. Together, this greatly increased $\Gamma_1$-$\Gamma_2$ separation towards 0° and 60° is indicative of increased interlayer coupling which produces a larger bonding-antibonding splitting of the $d_z{}^2$-manifold: For 0° and 60°, the AA or AB stacking provides a more direct hopping channel for the W atoms lying in each monolayer due to their alignment along the c-axis. This thus increases $d_{z2}$-$d_{z2}$ orbital overlap relative to lower-symmetry configurations. Note that cross-BZ moiré-driven hybridization with K point bands derived from in-plane orbitals would also hinder the interlayer hopping channels and reduce this separation. Analogous behaviour should be expected for chalcogen $p_z$ orbital interlayer hopping channels, relevant for the electronic structure at deeper binding energies.

Taken together, the energetic separation between the Γ and K point valence band maxima is minimized in the lowest-energy 0° and 60° configurations. In the intermediate regime, the separation is increased by ~100 meV for 30° misalignment. These changes can be explained entirely by the altering efficiency of interlayer hopping between W atoms localized in each monolayer. The insensitivity of the bands localized to the K points is indicative of negligible moiré-induced BZ reconstruction and cross-BZ hybridization, in-line with the apparent rarity of moiré-reconstructed bands away from specific angle regimes [8–13,48].

To finalize, we discuss the impact on the coupling to phonons within these twisted bilayer systems. It was previously observed in the monolayer limit how the proximity of the lowest energy bands in 1H-TMDs strongly influences the phonon coupling channels between the spin-split K-point valleys and the spin-degenerate Γ point [27]. $K_2$, which falls entirely below the local Γ point VBM, has several available scattering channels in momentum space. These channels are not available for band $K_1$ which lies closer to the Fermi level and above the Γ point VBM, reducing the electron-phonon coupling constant by two orders of magnitude. Since the K points are spin-polarized in the monolayer limit due to an absence of layer degeneracy, K to K' scattering is forbidden, and the efficiency of electron-phonon coupling is thus strongly spin dependent [27].

A similar, but more complex dependency should be present in twisted bilayers. As a function of twist angle, not only do the relative separations of these (now-)four bands change, but the degree of spin-polarization at the K point will also change due to the shifting alignment of the in-plane electric dipoles localized to each monolayer. The periodic dependence of the strength of the electron-phonon (el-ph) coupling with twist angle will not be in phase with the spin-dependency, however. The efficiency of the el-ph coupling channels from $K_1$ to $\Gamma_1$ will anticorrelate with the $K_1$-$\Gamma_1$ separation, thus finding a minimum at 30°. Simultaneously, scattering channels from $K_1$ to $K_1$' and $K_2$ to $K_2$' become increasingly inefficient as the relative twist angle approaches 60°. At 60°, as in free-standing monolayers, $K_1$ and $K_2$ are fully spin-polarized, maximising the imbalance between $K_1$ and $K_2$ of available scattering channels due to their energetic positioning either side of $\Gamma_1$. It is noteworthy that there will be an added dependence with the real space positioning on the heterostructure: As the relative orientation of W atoms across the van der Waals gap changes with lateral position on the sample, the relative band positions may be modified, thus altering the el-ph scattering efficiencies.

A similar dependence on twist angle should apply for other 2H structured TMDs, with potentially enhanced functional properties. Indeed, a reorganisation of the Γ and K point valence band maxima has been demonstrated in the 0 twist angle bilayer alloys $WS_{2(1-x)}Se_{2x}$ resulting in anomalously strong photoluminescence signals when numerous optical transitions coincide [49]. We note that all findings in Figure 3 are entirely in line with previous reports on twisted bilayer $MoS_2$ where similar changes to the $K_1$-$\Gamma_1$ separation were observed [50], though the much-reduced spin-orbit-driven $K_1$-$K_2$ separation will likely reduce the difference in el-ph coupling constants between $K_1$ and $K_2$, compared with monolayer $WSe_2$ [27] and discussed here for twisted bilayers.

## IV. CONCLUSIONS

Overall, from a systematic twist-angle dependent ARPES study of bilayer $WSe_2$ devices, we have quantified the energetic positions of the four lowest-energy valence bands with twist angle. The pair of K-point valleys have negligible twist angle dependence, reflecting an insensitivity to the details of the available interlayer hopping channels and therefore in limited moiré-driven cross-BZ hybridization. In contrast, the separation between the pair of quantized $k_z$-subbands at the Γ point reduces away from the high symmetry 0° and 60° configurations, in line with the hindered W-W hopping channels relative to AA/AB-stacked geometries. Together, our work highlights the ability to fine-tune the low energy energetics of twisted TMD heterostructures.

**ACKNOWLEDGEMENTS.**

M.T.E., M.S.F., O.J.C., T.-H.-Y.V., K.X. acknowledge funding support from ARC Discovery Project Grant No. DP200101345.

M.T.E. acknowledges funding support from ARC Future Fellowship FT2201000290 K.W. and T.T. acknowledge support from the JSPS KAKENHI (Grant Numbers 21H05233 and 23H02052) and World Premier International Research Center Initiative (WPI), MEXT, Japan. Support for crystal growth and characterization was provided by the National Science Foundation through the Penn State 2D Crystal Consortium-Materials Innovation Platform (2DCC-MIP) under NSF Cooperative Agreements No. DMR-2039351.

I.D.B. acknowledges support from the Ramón y Cajal program, grant no. RYC2022-035562-I.

J.B. acknowledges this research was supported by an AINSE Ltd. Postgraduate Research Award (PGRA).

This work was performed in part at the Melbourne Centre for Nanofabrication (MCN) in the Victorian Node of the Australian National Fabrication Facility (ANFF).

T.-H.-Y. Vu and O. J. Clark acknowledge travel funding provided by the International Synchrotron Access Program (ISAP) managed by the Australian Synchrotron, part of ANSTO, and funded by the Australian Government.

A. Grubišić-Čabo acknowledges the financial support of the Zernike Institute for Advanced Materials.

This research used resources of the Advanced Light Source, which is a DOE Office and Science User Facility under contract no. DE-AC02-05CH11231.

**Author Contributions**

T.-H.-Y.V. fabricated the $WSe_2$ devices with support from K.X. and W.Z.. Second Harmonic Generation was provided by J.B.M., L.J., and J.A.D.. S.H.L., Z.M., K.W., T.T. grew the crystals. T.-H.-Y.V., O.J.C., N.H.J., J.B., Q.L., I.D.B., A.G.C and M.T.E. performed the ARPES experiments with user support from S.H.R., C.J., A.B. and E.R.. T.-H.-Y.V., O.J.C. and M.T.E. wrote the manuscript with input from all co-authors. M.T.E. conceived the project and was responsible for overall project planning and direction.

**Data Availability**

Data are available from the authors upon reasonable request.

# Supplementary Information for: Tuning low-energy band structure in twisted bilayer $WSe_2$

T.-H.-Y. Vu[1,*], O. J. Clark[1,*], N. H. Jo[2], J. Blyth[1], Q. Li[1], C. Jozwiak[2], A. Bostwick[2], J. B. Muir[3,4], L. Jia[3,4], J. A. Davis[3,4], I. Di Bernardo[1,5], A. Grubišić-Čabo[6], K. Xing[1], W. Zhao[7,8], S. H. Ryu[2], S. H. Lee[9,10], Z. Mao[9,10], K. Watanabe[11], T. Taniguchi[12], B. A. Chambers[13], S. L. Harmer[13,14], E. Rotenberg[2], M. S. Fuhrer[1,8], M. T. Edmonds[1,8,15,†]

[1]School of Physics and Astronomy, Monash University, Clayton, VIC, Australia

[2]Advanced Light Source, Lawrence Berkeley National Laboratory, Berkeley, CA, 94720 USA

[3]Optical Sciences Centre, Swinburne University of Technology, Hawthorn, 3122, Victoria, Australia

[4]ARC Centre of Excellence in Future Low-Energy Electronics Technologies, Swinburne University of Technology, Hawthorn, 3122, Victoria, Australia

[5]Departamento de Física de la Materia Condensada, Universidad Autónoma de Madrid, Madrid, 28049 Spain

[6]Zernike Institute for Advanced Materials, University of Groningen, Nijenborgh 3, 9747 AG Groningen, The Netherlands

[7]Department of Materials Science and Engineering, Monash University, Clayton VIC 3800, Australia

[8]ARC Centre for Future Low Energy Electronics Technologies, Monash University, Clayton, VIC, Australia

[9]Department of Physics, Pennsylvania State University, University Park, PA, 16802, USA

[10]2D Crystal Consortium, Materials Research Institute, Pennsylvania State University, University Park, PA, 16802, USA

[11]Research Center for Electronic and Optical Materials, National Institute for Materials Science, 1-1 Namiki, Tsukuba 305-0044, Japan

[12]Research Center for Materials Nanoarchitectonics, National Institute for Materials Science, 1-1 Namiki, Tsukuba 305-0044, Japan

[13]Flinders Microscopy and Microanalysis, Flinders University, Bedford Park, South Australia 5042, Australia

[14]Institute for Nanoscale Science and Technology, Flinders University, Bedford Park, South Australia 5042, Australia

[15]ANFF-VIC Technology Fellow, Melbourne Centre for Nanofabrication, Victorian Node of the Australian National Fabrication Facility, Clayton, VIC 3168, Australia

* Equal contribution

† Corresponding author

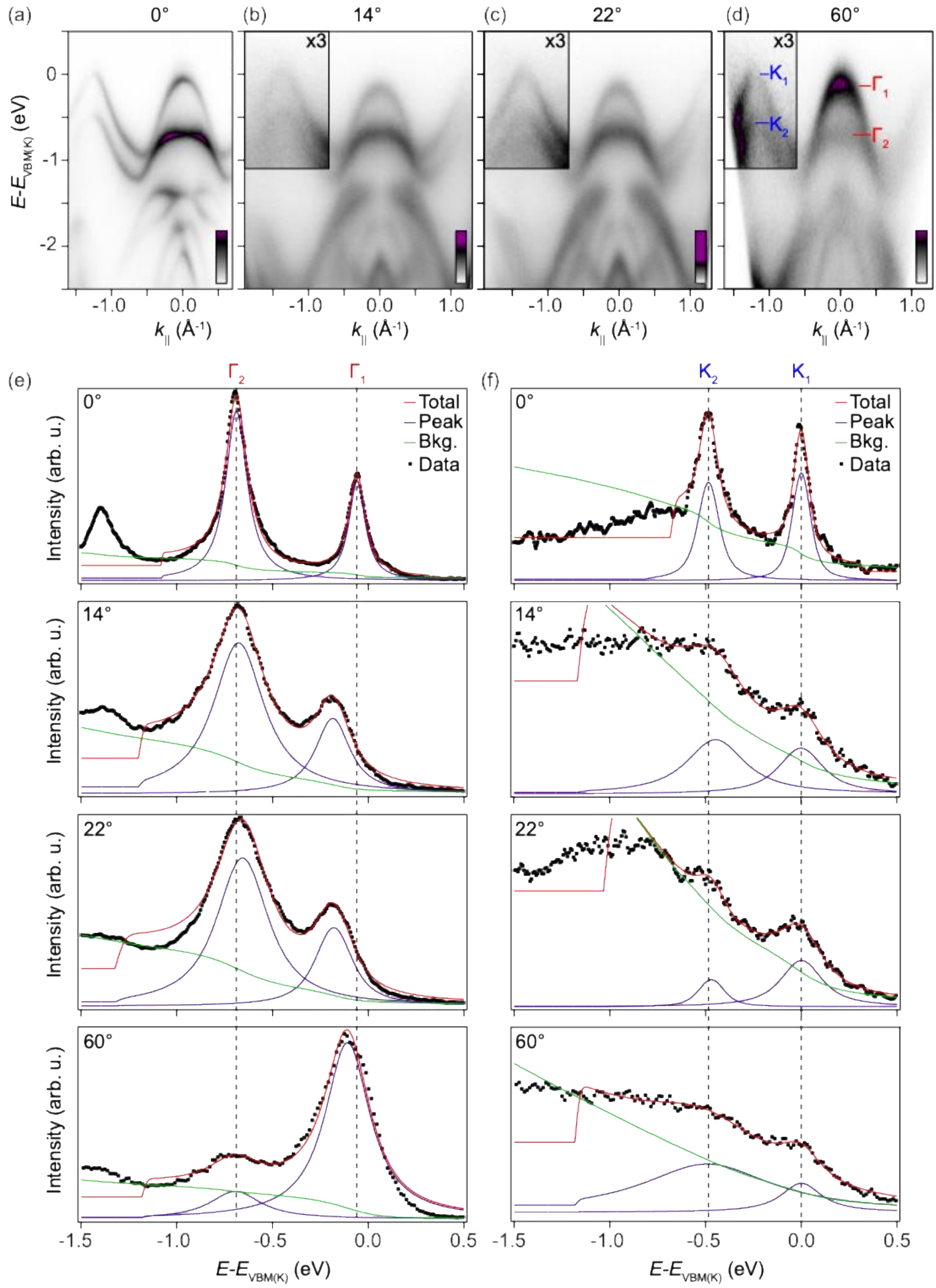

Figure S1 – Extracting the band positions as a function of twist angle. (a-d) Band dispersions reproduced from Figure 3(a-d) of the main text, without normalizations applied across the images. Regions of energy-momentum space within the black boxes have an increased colour contrast by a factor of 3 relative to the rest of the image. The energy axis is displayed relative to the K point valence band maxima ($K_1$) to remove uncertainties surrounding Fermi level positioning. In (d), the four highest energy bands are labelled for comparison with (e-f). (e-f) Energy distribution curves (EDCs) through the Γ (e) and K (f) points. The results of the fitting procedures are overlaid on the raw data. Vertical dashed lines in (e-f) are visual aids to show the changing positions of the four peaks as a function of twist angle.